\documentclass[%
 reprint,amsmath,amssymb,aps,
floatfix,
]{revtex4-1}
\usepackage{color}
\usepackage{graphicx}
\usepackage{dcolumn}
\usepackage{bm}

\begin{document}

\preprint{APS/123-QED}

\title{Strain-induced electronic phase transition and strong enhancement of thermopower of TiS$_{2}$}

\author{Atanu Samanta, Tribhuwan Pandey, and Abhishek K. Singh}
\affiliation{Materials Research Centre, Indian Institute of Science, Bangalore 560012, India}

\date{\today}

\begin{abstract}
Using first principles density functional theory calculations, we show a semimetal to semiconducting electronic phase transition for bulk TiS$_{2}$ by applying uniform biaxial tensile strain. This electronic phase transition is triggered by charge transfer from Ti to S, which eventually reduces the overlap between Ti-$(d)$ and S-$(p)$ orbitals. The electronic transport calculations show a large anisotropy in electrical conductivity and thermopower, which is due to the difference in the effective masses along the in-plane and out of plane directions. Strain induced opening of band gap together with changes in dispersion of bands lead to three-fold enhancement in thermopower for both \textit{p}- and \textit{n}-type TiS$_{2}$. We further demonstrate that the uniform tensile strain, which enhances the thermoelectric performance, can be achieved by doping TiS$_{2}$ with larger iso-electronic elements such as Zr or Hf at Ti sites.

\begin{description}
\item[PACS numbers]

\end{description}
\end{abstract}

\pacs{Valid PACS appear here}
                         
\maketitle

\section{Intorduction}
Transition-metal dichalcogenides (TMDs) are layered materials, which are well known for their unique electronic and optical properties~\cite{wang2012,Wilson1969}.  While within a layer, the metals and chalcogens form strong ionic-covalent bonds, these layers are held together by weak van der Waals (vdW) interactions. Depending upon the composition, TMDs offer a wide range of functional materials such as metals~\cite{Ayari2007}, semimetals~\cite{reshak2003}, semiconductors~\cite{Kam1982}, insulators~\cite{Sipos2008}, and superconductors~\cite{Morosan2006}. Among the TMDs, TiS$_2$ has been in focus of extensive research due to its potential applications as cathode materials for lithium-ion batteries~\cite{Friend1987,Whittingham1978,Mao1993,reshak2008}. Previous experimental studies~\cite{Imai2001} have reported that nearly stoichiometric TiS$_2$ shows a large power factor value of $37.1 ~\mu$W/K$^{2}$-cm at room temperature that is comparable with the best thermoelectric material Bi$_2$Te$_3$~\cite{HBTE}. The large power factor originates from a sharp increase in the density of states just above the Fermi energy as well as the inter-valley scattering of charge carriers~\cite{Friend1981,Koyano1986,Imai2001}. However, the semimetallic nature of TiS$_{2}$ gives rise to bipolar effects which are not desirable for thermoelectric applications~\cite{wood1988}. 

The electronic structure of TiS$_{2}$ is very unique and even a slight change can significantly influence thermoelectric properties. It has been shown that 0.04\% Mg doping at Ti-site in  TiS$_{2}$ causes a 1.6 times increase in thermopower at 300 K~\cite{Qin2007}. Recently, it has been discovered that the electronic structures of semiconducting TMDs (MX$_2$(M = Mo, W; X=S, Se, Te)) are very sensitive to the applied pressure/strain, which causes an electronic phase transition from semiconductor to metal~\cite{Dave2004,Bhattacharyya2012,Scalise2012,Yun2012}. Also few layers and mono-layer  TMDs show wide range of tunability in electronic and magnetic properties by application of strain~\cite{Zhou2012,Ma2012,Johari2012,Li2012,Shi2013,Shi2013}. Contrary to that, TiS$_{2}$ remains semimetallic up to a compressive hydrostatic pressure of 20 GPa~\cite{Bao2011}. So far there has been no studies on effect of uniform tensile strain on the electronic properties of TiS$_{2}$. Here, we carry out a study of  the electronic structure of TiS$_2$ as a function of applied uniform biaxial tensile strain (BTS). The material undergoes electronic phase transition from semimetal to a small band gap ($<0.15$ eV) semiconductor.  Most interestingly, the semiconducting strained TiS$_2$ exhibits nearly a four-fold enhancement in thermopower compared to the unstrained phases. We also explore the possibility of generating such a strain by doping with larger atoms at Ti sites. Iso-electronic Hf and Zr turn out to be the best dopants to generate the $2\%$ tensile strain producing the same enhanced thermoelectric properties as obtained by straining TiS$_{2}$.
\section{Computational Details}
Structural optimization, total energy and electronic structure calculations were performed using first principles density functional theory (DFT). The ionic cores are described by all-electron projector augmented wave potentials~\cite{PAW1, PAW2} and the Perdew-Burke-Ernzerhof~\cite{PBE} generalized gradient approximation (GGA) to the electronic exchange and correlation as implemented in the Vienna \textit{ab initio} simulation package (VASP)~\cite{PAW1,PAW2,Kresse}. We have incorporated the vdW interaction using Grimme's DFT-D2 method~\cite{Grimme2006,PAW1,PAW2,Kresse}(where D2 refers to the second generation of this method) for all the calculations. In this approach, the dispersion energy is described by an atom-pairwise interaction potential $E_{disp} = -\frac{1}{2}\sum\limits_{i\neq j}f_{damp}(R_{i,j})\frac{C_{6}^{i,j}}{R^{6}_{i,j}}$, where $R_{i,j}$ and $C_{6}^{i,j}$ are the distance and isotropic dipole-dipole dispersion coefficient between pair of atoms, respectively~\cite{Grimme2006}. The accuracy of the results crucially depends on the approximations of the dispersion coefficients and on the cutoff of damping function($f_{damp}$) at short distances ~\cite{Grimme2006,Tkatchenko2009,dftd22012,dftd22013}. The parameters used in the DFT-D2 method are taken from the empirical force-field of Grimme~\cite{Grimme2006}. The lattice parameters obtained using this approach are in very good agreement with experiments for various materials ~\cite{ dftd22013,He2014}. However, this approach overestimates the binding/cohesive energy  for most of the cases including TMDs ~\cite{dftd22013,Bhattacharyya2012,He2014}. 

The unit cell was optimized using the conjugate gradient scheme until the forces on every atom were $\leq$ 10$^{-3}$eV/{\AA}. For transport calculations, Boltzmann transport theory~\cite{ashcroft,Ziman} was used, which enables calculation of the temperature and doping level dependent thermopower and other transport parameters from the electronic structure. 
All the transport properties were calculated within the constant scattering time approximation (CSTA)~\cite{ashcroft,Ziman}. The CSTA is based on the assumption that the scattering time does not vary strongly with energy. It also does not consider any assumptions on temperature and doping level dependence of the scattering time. Within the CSTA, the energy dependence of transport function is described through both the density of states and carrier velocity. 
In this theory, the motion of an electron is treated semi-classically, and its group velocity in a specific band is given by
\begin{equation}
 \nu_\alpha(i, \textbf{k}) = \frac{1}{\hbar} \frac{\partial\epsilon(i,\textbf{k})}{\partial \textbf{k}_\alpha}
\label{eq:1}
\end{equation}
where $\epsilon(i,\textbf{k})$ and $\textbf{k}_\alpha$ are the ${i^{th}}$ energy band and $\alpha^{th}$ component of wavevector $\textbf{k}$, respectively. 
From group velocity $\nu_\alpha(i, \textbf{k})$ the thermopower and electrical conductivity can be obtained as
\begin{equation}
S_{\alpha\beta}(T, \mu) = \frac{1}{eT} \frac{\int\nu_\alpha{(i,\textbf{k})}\nu_\beta{(i,\textbf{k})}(\epsilon-\mu)\bigg[-\frac{\partial{f_{\mu}}(T,\epsilon)}{\partial\epsilon}\bigg]d\epsilon}{\int \nu_\alpha{(i,\textbf{k})}\nu_\beta{(i,\textbf{k})}\bigg[-\frac{\partial f_{\mu}(T,\epsilon)}{\partial\epsilon}\bigg]d\epsilon}
\label{eq:2}
\end{equation}

\begin{equation}
 \frac{\sigma_{\alpha\beta}(T, \mu)}{\tau (i, \textbf{k})} = \frac{1}{V} \int e{^2}\nu_\alpha{(i,\textbf{k})}\nu_\beta{(i,\textbf{k})}\bigg[-\frac{\partial{f_{\mu}}(T,\epsilon)}{\partial\epsilon}\bigg]d\epsilon
\label{eq:3}
\end{equation}
where $e$, $T$, $V$, $\tau$ and ${f_{\mu}}$ are the electronic charge, temperature, volume of the unit cell, relaxation time and Fermi-Dirac distribution function, respectively. Brillouin zone was sampled by a well converged 35$\times$35$\times$21 Monkhorst-Pack k-mesh~\cite{pack}. Subsequently, the group velocities were obtained by Fourier interpolation~\cite{synder,fourier} of band energies on converged  denser \textbf{k}-grid. These values are used in eqs.~\ref{eq:2} and ~\ref{eq:3}, to calculate the transport properties as implemented in BoltzTraP code~\cite{Madsen}. This approach has been shown to provide a good estimate of thermopower as a function of temperature and carrier concentration in a variety of thermoelectric materials without any adjustable parameters ~\cite{Jodin2004,Singh_2012,Singh2011,Lee2011,Parker2013,Pandey2013,Parker_AgGaTe}. 

\begin{figure}[h!]
\includegraphics[width=0.95\columnwidth]{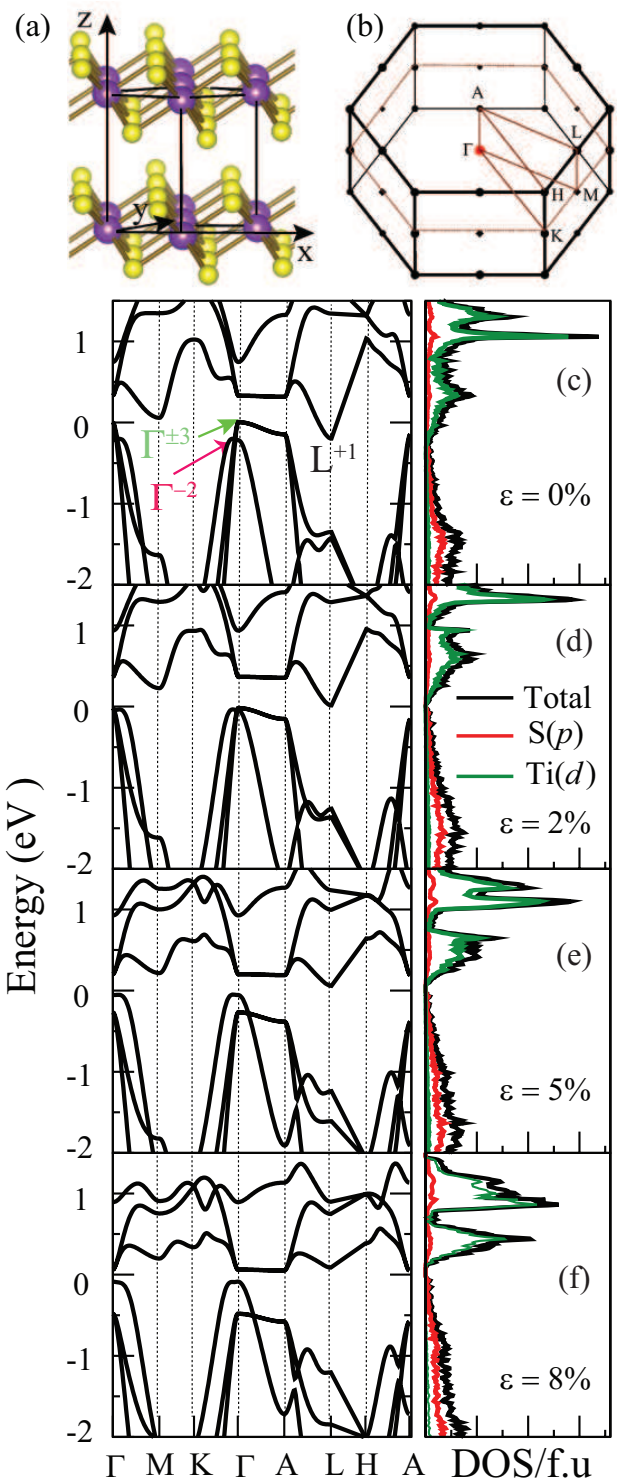} 
\caption{(a) Side view of the bulk unit cell of TiS$_2$ and (b) the corresponding Brillouin zone along with high symmetry points. Band structure, total and partial density of states per formula unit (f.u) at (c) 0 $\%$ (d) 2 $\%$ (e) 5 $\%$ and (f) 8 $\%$ strain values. Fermi energy is referenced to $0$ eV. } 
\label{fig:1} 
\end{figure} 
\section{Results and Discussion} 
TiS$_2$ crystallizes in a hexagonal structure with the space group P$\bar{3}$m1~\cite{Chianelli1975,Riekel1976}. The unit cell and corresponding Brillouin zone along with high symmetry points are shown in Fig.~\ref{fig:1}(a) and (b), respectively. The optimized lattice parameters a = b = 3.4 {\AA}, c = 5.77 {\AA} and  z = 0.246  are in good agreement with previously reported values~\cite{Chianelli1975,Riekel1976,Arnaud1981,Murray1972}. Calculated electronic band structure and corresponding  density of states (DOS) of bulk TiS$_2$ are shown in Fig.~\ref{fig:1}(c). The valence band maxima (VBM) and conduction band minima (CBM) at $\Gamma$-point originates from the S-\textit{(p)} and , Ti-\textit{(d)} orbitals, respectively. The overlap of these bands renders TiS$_{2}$ semimetallic. The extent of this indirect band overlapping is of 0.2 eV as shown in the Fig ~\ref{fig:1}(c). The valence bands $\Gamma^{-3}$/$\Gamma^{+3}$, and $\Gamma^{-2}$ mainly originate from S-(\textit{p}$_{x}$)/S-(\textit{p}$_{y}$), and S-(\textit{p}$_{z}$) orbitals~\cite{Benesh1985}, respectively. Among them, the first two bands are almost dispersion-less along $\Gamma$ to A in comparison to highly dispersive third band. This anomaly in dispersion comes from interlayer interaction of S-(\textit{p}$_{z}$) orbitals, which makes the third valence band more dispersive. The conduction bands at L (L$^{+1}$) and $\Gamma$ points originate from Ti-(\textit{d}$_{z}$) and Ti-(\textit{d}$_{x^{2}-y^{2}}$/\textit{d}$_{xy}$)~\cite{Benesh1985,shepherd1974,Mattheiss} orbitals, respectively. The conduction band along $\Gamma$ to A is less dispersive due to weaker interaction between two different layers.  On the other hand, due to stronger in-plane interaction of Ti-$d_{z}$ orbitals, the band at L point is highly dispersive. Therefore, the effective mass along $\Gamma$ to A is very high compared to the L symmetric points. This large difference in the effective mass makes the transport properties very anisotropic. 

\begin{figure}[!h]
\includegraphics[width=\columnwidth]{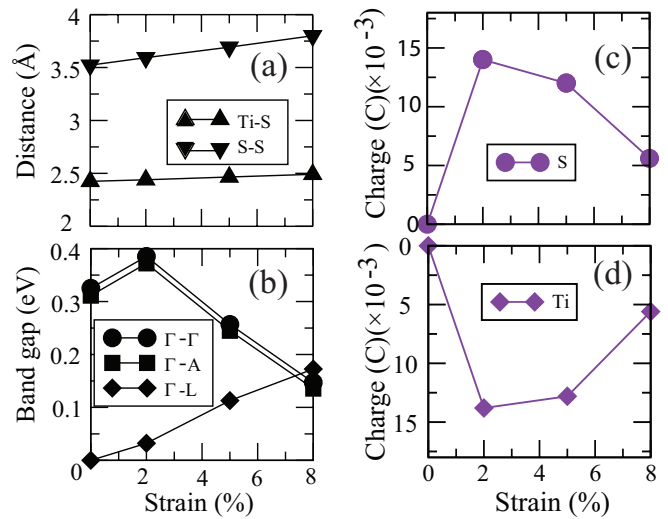}
\caption{(a) Change in bond length as a function of applied strain. (b) Change in band gap as a function of applied strain along different high symmetry directions. (c-d) Bader charge analysis as a function of strain.} 
\label{fig:2}
\end{figure}

It has been shown that strain can affect the electronic structure significantly by modifying the extent of the hybridization~\cite{Bhattacharyya2012}. As shown previously~\cite{Bao2011}, the compressive hydrostatic pressure does not affect the electronic phase of TiS$_{2}$. Therefore, we studied the effect of 2-8\% uniform BTS on electronic structure of TiS$_{2}$. The uniform BTS is applied along the x and y axis. Subsequently, each structure is relaxed at the fixed cell volume. The uniform BTS was calculated as $\epsilon$ = (a - a$_{0}$)/a$_{0}$ , where a$_{0}$ and a are the equilibrium and instantaneous lattice parameters, respectively. The corresponding band structure and DOS are shown in Fig.~\ref{fig:1}(d-f). The change in intralayer Ti-S and interlayer S-S distance with strain, is shown in Fig.~\ref{fig:2}(a). The intralayer bonding between S-(\textit{p})/Ti-(\textit{d}) orbitals is much stronger than the interaction between layers. Therefore, the S-S distance between two layers increases faster compared to the intralayer Ti-S distance with increasing strain, as shown in Fig.~\ref{fig:2}(a).

Next, we analyze the electronic band structures under the applied uniform BTS, which causes the electronic phase transition in this material. All the band gaps were calculated with respect to the
VBM at different strain values, irrespective of the symmetry of the VBM. The VBM remains at $\Gamma$
point for all the strain values. While the band gaps along $\Gamma-$A  and $\Gamma-\Gamma$ decrease, the gap along  $\Gamma-$L increases linearly with increasing strain as shown in the Fig.~\ref{fig:2}(b). At strain level lower than 7$\%$, TiS$_{2}$ has a indirect gap between the $\Gamma$ and the L points. When the strain exceeds the 7$\%$, the band gap shifts to $\Gamma-$A. With the increasing strain, the conduction band minima (CBM) at L becomes flatter due to decrease in the overlap of S-(\textit{p}) and Ti-(\textit{d}) orbitals, caused by increase in Ti-S bond lengths. Therefore, the bands around the Fermi energy in the conduction band become very heavy. The DOS at the Fermi-level in the conduction band also become sharp, as shown in Fig.~\ref{fig:1}(c-e). It is known that this type of 2D-nature of DOS enhances the thermoelectric performance, significantly~\cite{Parker2013,Hicks1993}. The third valance band along K-$\Gamma$-A moves up faster than the first and second valence bands, with increasing strain. This anomalous behaviour is caused by the faster increase in S-S interlayer distance than the S-Ti intralayer distance. Eventually, above the 8\% of applied strain the material becomes small-indirect to direct gap semiconductor. In order to gain further insight on semimetal-semiconductor transition, we compute the Bader~\cite{Henkelman2006,Tang2009} charge with FFT grid $400\times 400 \times 240$ and energy cutoff 1000 eV including all-electron charge (core + valence)~\cite{Aubert2011}, which are plotted in Fig.~\ref{fig:2}(c) and (d). The Bader charge analysis shows that with the applied strain charge transfer occurs between intralayer Ti to S atom thereby, increasing ionic bond character between Ti and S, which leads to the band gap opening.

\begin{figure}[h!]
\centering
\includegraphics[width=\columnwidth]{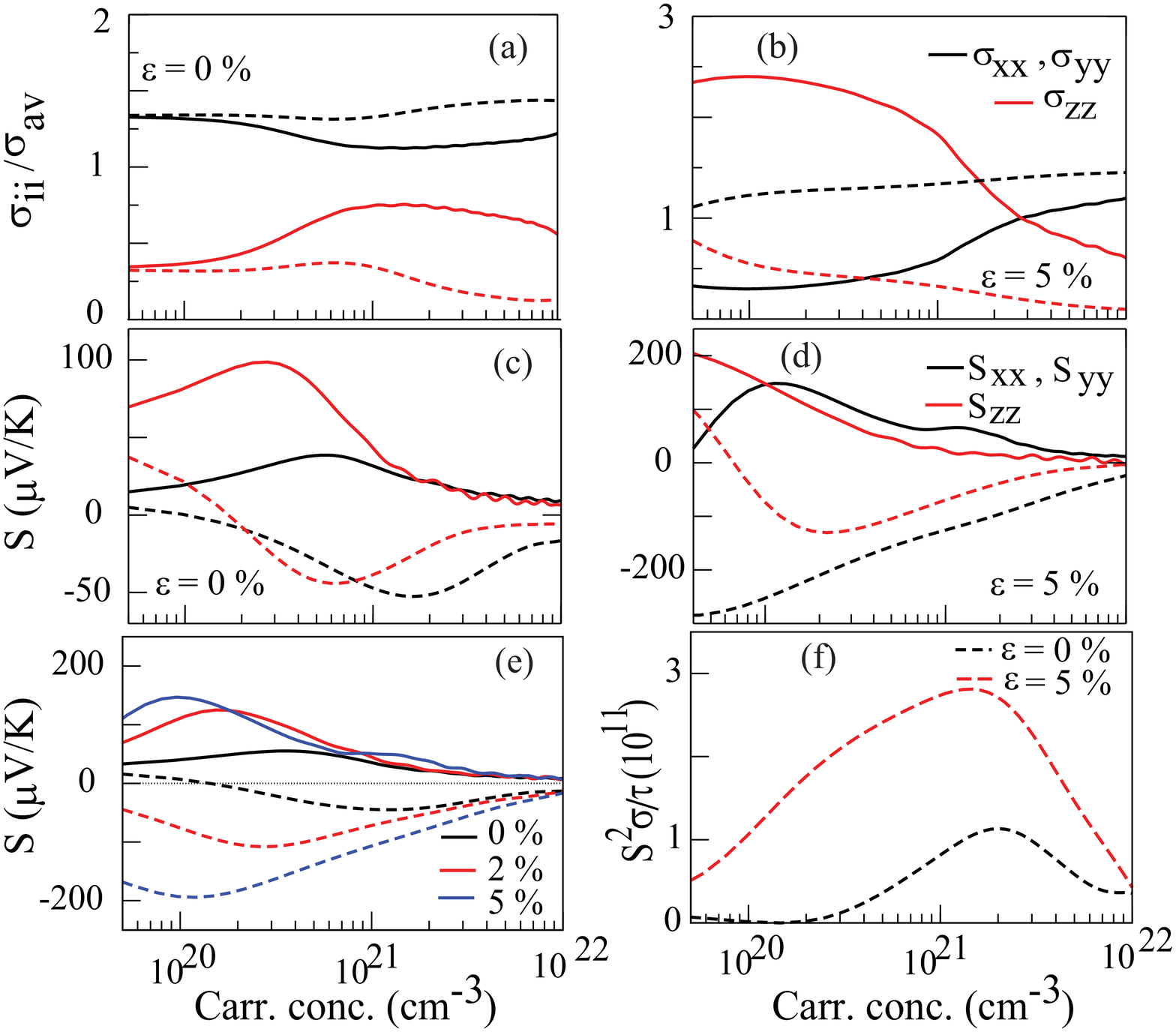}
\caption{(a) and (b)-Conductivity anisotropy ($\sigma_{ii}$/$\sigma_{av}$) as a function of carrier concentration for $p$-(solid lines) and $n$-type (dashed lines) TiS$_{\text2}$ at 0 and 5$\%$ strain, respectively. (c) and (d) Anisotropy in thermopower as a function of carrier concentration for $p$-(solid lines) and $n$-type (dashed lines) TiS$_{\text2}$ at 0 and 5$\%$ strain. (e) Dependence of average thermopower on the carrier concentration under different strain. (f) Power factor  divided by relaxation time ( in unit of W/m-K$^2$-s) as a function of carrier concentration. All these calculations are carried out at 300 K.}
\label{fig:3}
\end{figure}

It has been proposed that the semimetal to semiconductor transition can enhance thermoelectric performance~\cite{Dresselhaus2007,Heremans2000,Hicks1993}. Therefore, we study the effect of applied uniform BTS on the transport properties of TiS$_{2}$ within the constant scattering time approximation (CSTA)~\cite{ashcroft,Ziman}. The ratio of component of the conductivity tensor $\sigma_{ii}$ (where $i = x$, $y$, and $z$) to the average conductivity $\sigma_{av}$ (($\sigma_{xx}$+$\sigma_{yy}$+$\sigma_{zz}$)/3), which is independent of the relaxation time is calculated and shown in Fig.~\ref{fig:3}(a). The conductivity is the highest along the in-plane direction ($x$, $y$) for unstrained TiS$_{2}$ at 300 K over all the the carrier concentration range. Compared to in-plane directions, both conduction and valance bands along out of plane ($z$) direction ($\Gamma$-A in reciprocal space) have larger effective mass for electron and hole, leading to remarkable anisotropy in conductivity. While the hole conductivity increases significantly, with the applied uniform BTS, the electron conductivity remains nearly same as shown in Fig.~\ref{fig:3}(b). Interestingly, upon application of uniform BTS for \textit{p}-type TiS$_{\text2}$, the high conductivity direction changes from in-plane to out of plane in carrier concentration range of (5$\times$10$^{19}-$ 10$^{21}$ cm$^{-3}$).  At a very high carrier concentration (excess 5$\times$10$^{22}$ cm$^{-3}$) in-plane conduction starts to dominate again. On the other hand for \textit{n}-type TiS$_{\text2}$, the conduction direction remains in-plane over all the range of carrier concentration. However, the difference between in-plane and out of plane conductivity gets reduced compared with unstrained case. A similar behavior of conductivity has also been observed in other layered compounds such as unstrained CrSi$_{\text2}$ and FeSi$_{\text2}$ ~\cite{Singh_2012,Pandey2013}.

The reason for this dependence of anisotropy in conductivity on the applied uniform BTS can be found in the band structure as shown in Fig~\ref{fig:1}(c-d). At 0 $\%$ strain Fig~\ref{fig:1}(c), the valence band is flatter along $\Gamma$-A, leading to lower hole mobility. On the other hand the conduction band has two pockets at M and L symmetry points, leading to higher mobility for electrons, which is consistent with the conductivity behavior as shown in Fig.~\ref{fig:3}(a). With the increasing strain the third valence band, which is highly dispersive moves up and finally crosses the flatter band at around 5\% strain. This lowers the effective mass of the holes dramatically, leading to drastic increase in out of plane conductivity, which starts to dominate over the in-plane conductivity as shown in Fig.~\ref{fig:3}(b). On the other hand, with the applied uniform BTS the dispersion of the conduction band reduces, causing a decrease in the anisotropy in conductivity, as shown in Fig.~\ref{fig:3}(b). However, for a complete understanding of electrical conductivity, one needs to estimate the effect of different scattering mechanisms on relaxation time ($\tau$). The carrier transport in a system is mainly hampered by scattering of charge carriers due to phonons (acoustic, optical, and polar-optical) and point defects~\cite{Ravich_book,Ravich_paper,Ahmad2010}.  For non parabolic bands the average band mass (density-of-states effective mass for each pocket) is defined as $m_{b}=(m_{L}m_{T}^{2})^{1/3}$ where, $m_{L}$ and $m_{T}$ are longitudinal and transverse effective masses repectively~\cite{Ravich_book,Ahmad2010}. According to the deformation potential theory of Bardeen ~\cite{Bardeen} the relaxation time $\tau \propto m_{b}^{-3/2}$~\cite{Bardeen,Herring1956,Ziman,Ahmad2010}. The band mass of the charge carriers increases with increasing strain, which leads to the decrease in the relaxation time as well as conductivity. At low temperatures, charge carriers are scattered mostly by the point defects ~\cite{Ravich_book,Ravich_paper}, which can further reduce the relaxation time. All the scattering mechanisms described above will affect the transport properties. Furthermore, for chalogenides at high temperatures, the scattering is mainly dominated by phonons~\cite{Ravich_book,Ahmad2010}. Since our calculation does not involve explicitly the effect of these scattering mechanisms, the results present the upper-limit of the transport properties. However, it is shown experimentally that irrespective of all these scattering effects, TiS$_{2}$ still shows good electrical conductivity~\cite{Imai2001,TiS2_cond}. Thereby, CSTA can be good approximation for TiS$_{2}$.

In order to estimate the effect of uniform BTS on thermoelectric performance of TiS$_{2}$, we have calculated the thermopower of the unstrained and strained TiS$_{2}$. The direction dependent thermopower  at 300 K as a function of carrier concentration for 0 and 5 $\%$ strain is shown in Fig.~\ref{fig:3}(c) and (d), respectively. Like conductivity, the thermopower also exhibits large anisotropy at low concentration for both \textit{p}- and \textit{n}-type carriers. However, at high carrier concentration the thermopower shows reduced anisotropy. This anisotropy comes from the non parabolic nature of bands which gets modified by applying strain. The density of states at $5\%$ strain becomes sharp near the Fermi energy at the conduction band as shown in the Fig.~\ref{fig:1}(e), which results in large thermopower for a given carrier concentration. Furthermore, the band structure of TiS$_{2}$ is composed of heavy and light mass bands near the Fermi energy that also contributes to high thermopower values. A similar phenomenon has also been observed for many other good thermoelectric materials~\cite{Parker_AgGaTe,Singh2011,Singh_2012}. 

At low carrier concentration and 0 $\%$ strain, the thermopower attains high values along out of plane for both \textit{p}-type and \textit{n}-type cases. The thermopower increases with the applied strain attaining a maxima at 5 $\%$ strain. With the strain TiS$_{2}$ shows large thermopower for both \textit{p}-type and \textit{n}-type doping along out of plane and in-plane direction, respectively. This improved performance comes from opening of band gap and changes in the band dispersion with the applied strain.

Next, we study the effect of applied uniform BTS on the average of all the components of thermopower, as it enables the comparison of our results with the polycrystalline samples. The average thermopower is defined as $1/3$ of trace of the thermopower tensor ($S_{av} = (S_{xx}+S_{yy}+S_{zz})/3$). The average thermopower as a function of carrier concentration shows a peak, which shifts with the applied strain as shown in Fig~\ref{fig:3}(e). This peak arises due to bipolar effect. For a good thermoelectric performance a material should exhibit unipolar behavior over a wide range of carrier concentration \cite{wood1988}. As shown in Fig~\ref{fig:3}(e), the calculated averaged thermopower for \textit{p}- and \textit{n}-type at 300K, show unipolar behavior for a large range of carrier concentration compared to unstrained case. For \textit{p}-type TiS$_{2}$ at 5 $\%$ strain  highest thermopower can be achieved at moderate carrier concentrations typically, $1-5$ $\times$10${^{20}}$ cm${^{-3}}$. We report for \textit{n}-type TiS$_{\text2}$ efficient thermoelectric performance can be achieved in the carrier concentration range of $1-7$ $\times$10${^{20}}$  cm${^{-3}}$ depending upon temperature. The above predicted doping level is equivalent to approximately $0.006-0.03$ holes and $0.006-0.04$ electrons per unit cell, respectively. Previous experimental study of Sams \textit{et. al.} ~\cite{doping2005} has demonstrated that TiS$_{\text2}$ can be heavily doped substitutionally at both Ti and S sites. Based on this experimental observation it can be assumed that doping level proposed here should be readily achievable.

As discussed above due to the presence of heavy band and multi-carrier transport, \textit{n}-type leads to higher thermopower than \textit{p}-type even at higher carrier concentrations.  With the applied strain, the thermopower increases nearly four times in comparison to unstrained case, for the both type of the carriers. However, strain not only affects the thermopower but also the electrical conductivity ($\sigma$). In order to quantify the effect of strain on thermoelectric performance ($ZT = S^{2} \sigma/ \kappa_{total}$, where $\kappa_{total}$ is the total thermal conductivity) it is important to calculate power factor ($S^2\sigma$). The power factor is defined as the ability of a material to produce useful electrical power at a given temperature difference. The large power factor is indicative of better thermoelectric performance.  As a first approximation for power factor ($S^2\sigma$), here the quantity $S^2\sigma/\tau$ is calculated at 300K for \textit{n}-type TiS$_{2}$ as shown in Fig.~\ref{fig:3}(f). This quantity also increases drastically (three times) in comparison to the unstrained case. The figure of merit ZT closely follows the overall change in the thermoelectric performance hence, can also be enhanced by three times assuming no drastic changes in thermal conductivity and relaxation time.

It is well known that LDA/GGA tends to underestimate the band gap~\cite{Sham1983, Bhattacharyya2012}. In order to test the effect of band gap underestimation on thermoelectric performance, we applied the scissors operator and adjusted the band gap to 0.30 eV for 5$\%$ strained TiS$_{2}$. The scissors operator was applied by rigidly shifting the conduction bands up and the valence bands down by 0.075 eV around the mid gap energy. After the rigid shift of the bands we performed the transport calculations. The maximum value of $S^2\sigma/\tau$ is 8 $\times$ 10$^{11}$ W/m-K$^2$-s, which is nearly twice than the value at the band gap 0.15 eV. Therefore, the GGA results serve as a lower limit for the thermoelectric performance. 

\begin{figure}[!htbp]
\centering
\includegraphics[width=\columnwidth]{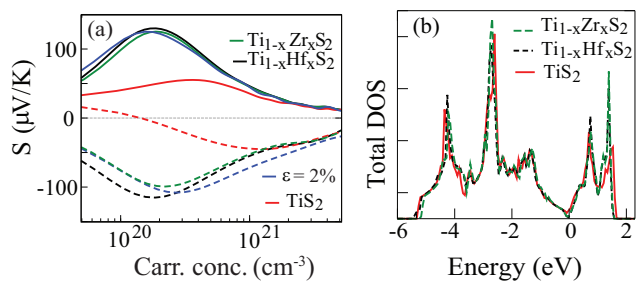}
\caption{(a) The average thermopower as a function of the carrier concentration for \textit{p}-(solid lines) and \textit{n}-type (dashed lines) at 300 K (b) Total density of state of Hf ($x$ = 0.25) and Zr ($x$ = 0.25) doped TiS$_{2}$, respectively. For comparison purpose the DOS of bulk TiS$_{2}$ is also shown in the same figure. }
\label{fig:4}
\end{figure}

Having shown the advantages of applied strain on the electronic and thermoelectric properties of TiS$_{2}$, next we look at the feasibility of generating such kind of strain in experiments. In order to achieve uniform BTS intrinsically in bulk TiS$_{2}$, the material can be grown epitaxially on a substrate with larger lattice parameter. However, the presence of interface would dominate the electronic properties, masking the intrinsic changes in properties by BTS.  Another more easier route could be doping with the iso-electronic atoms of larger size, which could induce the required BTS. We have studied doping of TiS$_{2}$ using a $2\times2\times2$ supercell with the iso-electronic Hf and Zr on Ti site. The volume relaxation of the doped structures were carried out by including the vdW interaction between the layers. The fully relaxed structure expands due to larger size of the Zr and Hf. This eventually generates in-plane tensile strain of the order of $2\%$ and $1.98\%$  corresponding to a doping of  0.25 Zr and Hf atom per formula unit, respectively. The out of plane strain is nearly zero for both the dopants. The calculated thermopower of iso-electronic doped systems matches very well with that of $2\%$ tensile strained bulk TiS$_{2}$ for both \textit{p} and \textit{n}-type doping, as shown in Fig.~\ref{fig:4}(a). The DOS (Fig.~\ref{fig:4}(b)) does not show any significant changes far away from Fermi energy with iso-electronic doping. However, the reduction in DOS around the Fermi energy is similar to the one caused by $2\%$ tensile strain. 
For TiS$_{2}$, better thermoelectric performance can be achieved at optimized strains ($2-5\%$) as well as optimized carrier concentrations (approximately, $0.006-0.03$ holes and $0.006-0.04$ electrons per unit cell). As shown experimentally ~\cite{doping2005}, TiS$_{2}$ can be heavily doped with V, Mo, Ta and Ce at Ti-site. Since, the ionic radii of these dopants are larger than the Ti-ionic radius, these dopants can generate lattice strain and charge carriers in the system. For practical realization of strain values proposed here, one can perform co-doping by isovalent and  \textit{p}- or \textit{n}-type dopants, having larger ionic radius than Ti.
\begin{table}[!ht]
\centering
\begin{tabular}{|c|c|c|c|c|}
\hline 
\multicolumn{1}{|c|}{System}& \multicolumn{1}{c|}{Direction} & \multicolumn{3}{c|}{Sound velocity(m/sec)} \\[0.9ex]
 \cline{3-5}
\multicolumn{1}{|c|}{} & \multicolumn{1}{c|}{}  & \multicolumn{1}{c|}{$v_{TA1}$} & \multicolumn{1}{c|}{$v_{TA2}$} & \multicolumn{1}{c|}{$v_{LA}$}\\[0.9ex]
\hline
   TiS${_2}$ & $\Gamma$-M  & 2510.54 & 3693.01 & 5332.23 \\[1.5ex] 
& $\Gamma$-A  & 2000.64 & 2000.64 & 3735.94 \\[1.5ex] 
  \cline{3-5}
  \cline{0-0}
  \textbf{Average}  & & 2255.59 & 2846.82 & 4534.08\\[1.5ex] 
\hline    
  Ti$_{0.94}$Zr$_{0.06}$S$_{_2}$ & $\Gamma$-M   & 2247.00 & 2856.00 & 3587.00 \\[1.5ex] 
  & $\Gamma$-A   & 2070.33 & 2070.33  & 3542.00 \\[1.5ex] 
  \cline{3-5}
  \cline{0-0}
  \textbf{Average} & & 2158.66 & 2463.16 &3564.50 \\[1.5ex] 
  \hline
   Ti$_{0.83}$Zr$_{0.17}$S$_{_2}$ & $\Gamma$-M   &  1964.02 & 1643.40 & 1893.21 \\[1.5ex] 
    & $\Gamma$-A   &  1925.02 & 2058.09 & 3362.30 \\[1.5ex] 
  \cline{3-5}
  \cline{0-0}
  \textbf{Average} & & 1944.52 & 1850.74 & 2627.75\\[1.5ex]
\hline 
\end{tabular}
\label{Table:1}
\caption{Effect of doping on transverse $v_{Trans}$, longitudinal $v_{Long}$ and average sound velocities. Sound speeds in m/s are extracted from phonon dispersion data. $v_{Trans}$ and $v_{Long}$ denotes speed corresponding to transverse and longitudinal modes respectively.}
\end{table}

\begin{figure}[!htbp]
\centering
\includegraphics[width=\columnwidth]{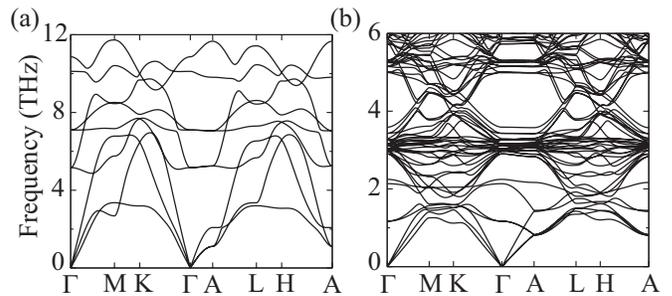}
\caption{Phonon dispersion (a) bulk TiS$_{2}$ and (b) Ti$_{0.94}$Zr$_{0.06}$S$_{_2}$.}
\label{fig:5}
\end{figure}

Lattice part of thermal conductivity $\kappa_{l}$ is an another key parameter, which influences the thermoelectric performance of a material. Zr and Hf turn out to be the best iso-electronic dopant, which enhance the thermoelectric performance of TiS$_{2}$. In order to get qualitative estimate of thermal conductivity of Zr doped TiS$_{2}$, we have calculated phonon dispersion of undoped TiS$_{2}$ and doped Ti$_{1-x}$Zr$_{x}$S$_{2}$ $(x= 0.06$ and $0.17$), which are shown in Fig. ~\ref{fig:5} (a) and (b). The density functional perturbation theory(DFPT) in VASP is used to calculated force constants for phonon calculation. Then, the phonon dispersions are calculated using Parlinski-Li-Kawazoe method as implemented in PHONOPY package~\cite{phonopy}. 
The phonon group velocity is calculated from the relation $v_{g}= \frac{\partial \omega}{\partial q}$ as shown in Table~\ref{Table:1}, where $q$ is the wave vector. According to the single mode relaxation time approximation of the Boltzmann equation, the total thermal conductivity ($\kappa_{i}$) contributed by the each phonon mode  can be represented as $\kappa_{i}(q)= C_{i}(q) v^{2}_{i}(q) \tau_{i}(q)$, where $C_{i}$, $v_{i}$, and $\tau_{i}$ are the specific heat,  the group velocity, and  the phonon relaxation time of $i^{\text{th}}$ mode, respectively. $\kappa_{l}$ is directly proportional to square of phonon group velocity, implying that decrease in phonon group velocity will reduce the lattice thermal conductivity. As presented in Table ~\ref{Table:1}, with the increasing Zr concentration the phonon group velocity significantly decreases along in-plane and out-of-plane direction, which will lower the thermal conductivity. This reduction in phonon group velocity is due to mass difference between Zr and Ti atoms. Furthermore, the large differences in atomic mass and size can increase the phonon scattering ~\cite{Takuma2014,Yan2012}, which will lead to further reduction of thermal conductivity. 

\begin{figure}[!htbp]
\centering
\includegraphics[width=0.75\columnwidth]{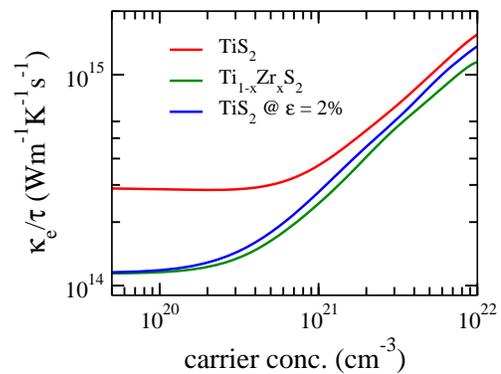}
\caption{Carrier concentration dependent electronic part of the thermal conductivity scaled with respected to relaxation time.
respectively }
\label{fig:6}
\end{figure}

Because of semi-metallic nature of TiS$_{2}$ the electronic thermal conductivity ($\kappa_{e}$) could hamper the thermoelectric performance. However, under strain TiS$_{2}$ exhibits a electronic phase transition from semi-metal to semiconductor which can lower the $\kappa_{e}$.  In Fig.\ref{fig:6} we present the $\kappa_{e}$  scaled with respect to $\tau$ for bulk, 2\% strain and Zr doped TiS$_{2}$. As can be seen at lower carrier concentration we observe a significant reduction in $\kappa_{e}$ for Zr doped TiS$_{2}$ and 2\% strained TiS$_{2}$ in comparison to bulk TiS$_{2}$. On the other hand, at higher carrier concentration the thermal conductivity remains more or less same for all the cases. Based on this analysis we can safely conclude that the total thermal conductivity ($\kappa_{e} + \kappa_{l}$) will decrease under doping. Thus, the observed enhancement in thermoelectric performance as shown in this work can be achieved experimentally by doping/co-doping of TiS$_{2}$.
\section{Conclusions}
In summary, we demonstrate in TiS$_{2}$, the electronic phase transition from semimetallic to semiconducting by application of uniform biaxial tensile strain. The strain modifies the dispersion of bands, which improves the thermoelectric performance of the material, significantly. Our results show a large strain dependent anisotropy in electrical conductivity and thermopower at lower carrier concentration for both \textit{p}- and \textit{n}-type carriers. The thermopower results suggest that 5 $\%$ strained TiS$_{\text2}$ exhibits three-times enhancement in thermoelectric properties. We further demonstrate the possibility of generating such strain by doping of TiS$_{2}$ with larger size atoms such as Zr and Hf at Ti sites. Moreover, experimental realization of strain, induced by doping can serve as a powerful tool to enhance the electronic and thermoelectric properties of this promising material.
\begin{center}                                  
\textbf{ACKNOWLEDGMENTS}
\end{center}
This work was supported by the ADA under NPMASS and DST nanomission. The authors thank the Supercomputer Education and Research Centre and Materials Research Centre, IISc, for providing the required computational facilities for the above work.


%

\end{document}